\documentclass[12pt]{article} 
\usepackage{epsfig}

\begin {document}
~\hfill Physics Letters B, revised submission 21/9/00\\[2cm]

\begin{center}
{\Large \bf An infrared background--TeV gamma-ray crisis?}\\[1cm]
R. J. Protheroe$^1$ and H. Meyer$^{1,2}$\\
$^1$Department of Physics and Mathematical Physics,\\
The University of Adelaide, Adelaide, Australia 5005.\\
$^2$Permanent address: Universit\"{a}t Wuppertal, Fachbereich Physik,\\ 
GaussStr 20, 42097 Wuppertal, Germany.\\
\end{center} 

\begin{abstract}
We consider the implications of the recent determination of the
universal infrared background for the propagation of photons up
to 20 TeV from the active galaxy Markarian 501 as observed by
HEGRA.  At 20 TeV the mean free path for photon-photon collisions
on the infrared background would be much shorter than the
distance to Markarian 501, implying absorption factors of the
order of $\exp (-10)$, or greater, and consequently an excessive
power output for this active galaxy.  Possible solutions of this
problem are discussed.\\
\end{abstract}


\section {Introduction}

The general form of the spectrum of the universal infrared background
radiation was predicted by Primack et al. \cite{Primack}, Malkan and
Stecker \cite{MalkanStecker}, Dwek et al. \cite{Dwek} and Fall et
al. \cite{Fall}.  The origin is currently thought to be due to star
formation in the early universe producing starlight predominantly in
the 1 micron range which was then re-processed into the 100 micron
range due to heating of surrounding dust.  Early data from the DIRBE
and FIRAS instruments on COBE (see ref.~\cite{Dwek} for a review) and
from ISO \cite{ISO} were mainly upper or lower limits which did not
severely constrain the models.  New results from these experiments are
becoming available due to a more sophisticated treatment of the
galactic and solar-system radiation foreground subtraction.  In
particular, new determinations of the infrared background intensity
have very recently become available at 1.25, 2.2 and 3.5 microns
\cite{Wright}, 15.0 microns \cite{Biviano}, 60 and 100 microns
\cite{Finkbeiner}, and 140 and 240 microns \cite{Hauser}.  In Fig.~1
we show the predictions of Malkan \& Stecker \cite{MalkanStecker}
together with the new observations and recent lower limits
\cite{WangBiermann} based on infrared galaxy source counts and a model
including the effects of an infrared burst phase of ultraluminous
infrared galaxies.  The observations are in excellent agreement with
prediction inasmuch as the expected trend (two peaks and valley) is
found, the agreement being particularly good between 1.25 and 15
microns. However, the level of the observed flux in the 60--240 micron
range is higher, and in the case that this is a truly universal
background it has important consequences both for star formation
in the early universe, and for intergalactic absorption of TeV
gamma-rays which leads to a crisis in our understanding of high energy
phenomena.

Immediately following the discovery of Markarian 421 as a TeV
source  \cite{Punch} it became clear that interactions with
a universal infrared background may have a strong influence on
the propagation of TeV gamma-rays through intergalactic space
 \cite{Stecker,ProtheroeStanev}, in the same way that the 
microwave background affects the propagation of 1000~TeV
gamma-rays  \cite{GouldSchreder,Jelley}.  The observation by
several telescopes of a very high level of emission by Markarian
501 during 1997 (see ref.~\cite{Protheroe97} for a review) showed
that the spectrum extended to 10 TeV and beyond.  At that time,
it was pointed out by Stanev and Franceschini
 \cite{StanevFranceschini} that if the spectrum continued much
beyond 10 TeV that we would ``have to revise our concepts about
the propagation of TeV gamma-rays in the intergalactic space,
and that something complicates the process.''  This has now come
about as the analysis of HEGRA data for the whole of 1997 shows
that the spectrum of Markarian 501 extends well beyond 10
TeV \cite{HEGRA97data501}.

\section {Implications of the new IR observations for the propagation 
of Gamma-rays through the universe}

Photon-photon pair production is a very elementary process in
QED, and can safely be calculated at energies relevant to TeV
astronomy.  For propagation of energetic particles (gamma-rays) of energy
$E_\gamma$, mass $m_\gamma$(=0) and velocity $\beta_\gamma c$(=$c)$,
through isotropic radiation the reciprocal of the mean
free path for collisions with photons is given by
\begin{equation}
x_{\gamma \gamma}(E_\gamma)^{-1} = {1 \over 8 {E_\gamma}^2\beta_\gamma}
\int_{\varepsilon_{\rm min}}^{\infty} \, d\varepsilon
\frac{n(\varepsilon)} {\varepsilon^2} \int_{s_{\rm min}}^{s_{\rm
max}(\varepsilon,E_\gamma)} ds \, (s - m_\gamma^2c^4) \sigma(s),
\label{eq:mpl}
\end{equation}
where $n(\varepsilon)$ is the differential photon number density,
and $\sigma(s)$ is the total cross section for a centre of momentum
frame energy squared given by $s=m_\gamma^2c^4 + 2 \varepsilon
E_\gamma(1 - \beta_\gamma\cos
\theta)$ where $\theta$ is the angle between the directions of
the energetic particle (gamma-ray) and soft photon, $s_{\rm min}
= (2 m_e c^2)^2$, $\varepsilon_{\rm min} = (s_{\rm
min}-m_\gamma^2c^4 ) / [2E_\gamma(1+\beta_\gamma)]$, and $s_{\rm
max}(\varepsilon,E_\gamma) = m_\gamma^2c^4 + 2 \varepsilon
E_\gamma(1 + \beta_\gamma)$. For photon-photon pair production by
gamma-rays, we take $m_\gamma=0$ and $\beta_\gamma=1$, and the
photon-photon pair production cross section from 
 \cite{JauchRohrlich55}.

We have calculated the mean free path for this process for
interactions of gamma-rays with the infrared and cosmic microwave
backgrounds, and this is shown in Fig.~2.  Also shown are results
\cite{BednarekProtheroe99} based on the upper and lower models
for the IR background due to Stecker and Malkan
\cite{MalkanStecker}, extended to shorter wavelengths.  We note
that above 10~TeV the mean free path using the new IR intensity
(thick solid curve) is almost an order of magnitude lower, and
this has very important consequences for the HEGRA observations
of Markarian 501.

We show in Fig.~3 the time-averaged flux from Markarian 501
observed by HEGRA using the stereoscopic system of telescopes
 \cite{HEGRA97data501}.  We note that the spectrum
smoothly extends to at least 20~TeV.  This observation was
confirmed by using the stand-alone telescopes of HEGRA, although
with less statistical significance and with rather different
systematics  \cite{HEGRA501standalone}.  It is important to note
that the Crab Nebula, as a steady galactic source, has
approximately the same emission level as Markarian 501 in the
high state, and has been continuously monitored by the HEGRA
telescopes  \cite{HEGRA501standalone,HEGRAcrab}.  The Crab
spectrum is also shown in Fig.~3 for comparison (shifted down by
a factor of 100 for clarity).  The Crab Nebula spectrum
determined extends again up to 20 TeV as a straight power-law
with a very similar spectral index, and this suggests reliable
detection of photons up to 20 TeV.  For the following discussion,
an understanding of the absolute energy calibration is of
considerable importance.  This was investigated in great detail
by Monte Carlo methods, and in addition, checked by absolute
measurement of the cosmic ray proton flux using the HEGRA
telescope system in about the same energy range as in the Crab
and Markarian 501 observations.  In this energy range the cosmic
ray proton flux is rather well measured by direct methods, and a
compilation of data can be found in
ref.~\cite{BiermannWiebel-Sooth}.  The absolute energy scale of
HEGRA was then estimated to have an uncertainty of 15\%
 \cite{HEGRAcalibration}.  In addition, there is data available on
Markarian 501 from the Whipple  \cite{Whipple501} and CAT
 \cite{CAT501} telescopes, and this is in very good agreement with
the HEGRA data but does not reach as high in energy, mainly
because of their lower exposure.

We have applied the mean free path in the infrared background
radiation plotted in Fig.~2 to ``correct'' the observed flux from
Markarian 501 assuming a Hubble constant of $H_0=65$
km~s$^{-1}$~Mpc$^{-1}$, and this is also shown in Fig.~3.  We
note that this implies a dramatic (by several orders of
magnitude) increase in luminosity at the source (see right hand
scale).  Even taking into account relativistic beaming, with a
typical Doppler factor of 20, this would imply an extraordinarily
high luminosity for an active galaxy of the BL Lac class of which
Markarian 501 is a member.  The large correction factor is a
direct consequence of the new IR data at 60 and 100 microns and
the implied shift of the ``dust peak'' to somewhat shorter
wavelengths.  It is worthwhile pointing out that the correction
factor depends on the value of the Hubble constant.  If it were
larger than that we have assumed, then the distance to the
source would be smaller, and the absorption less, but
unacceptably high values of $H_0$ would be required to give a
reasonable source luminosity.  In fact, the problem may be even
more severe if future infrared measurements would fill in the
``valley'' between 5 and 40 microns giving an IR background such
as shown by the short dashed lines in Figure 1.  In this case,
the mean free path would be as shown by the short dashed curves
in Figure 2, with the result that correction for absorption would
cause an even more dramatic rise of the source spectrum above a
few TeV.

\section {Immediate consequences} 

The source spectrum of Markarian 501, i.e. the observed spectrum
corrected for propagation from the source, is in gross
disagreement with expectations of the Synchrotron self-Compton
(SSC) model routinely used to interpret the observed spectra of
active galaxies.  In this model, accelerated electrons produce
most of the observed radio through X-ray emission as synchrotron
radiation, and the same population Compton scatters synchrotron
photons to gamma-ray energies.  The apparent success of the SSC
model depends critically on which intergalactic absorption is
used, as the observed sources are at the gamma-ray horizon
 \cite{beakons}.  Now that the most recent determinations of the
diffuse IR background are high, this makes it impossible to fit
the gamma-ray observations using the SSC model.  For example, Guy
et al. \cite{Guy} took a relatively high level for the IR
background, but still a factor of 2 short of recent
determination.  Petry et al. \cite{Petry} and Bednarek and
Protheroe  \cite{BednarekProtheroe99} used the IR background at a
level of that predicted by Malkan \& Stecker \cite{MalkanStecker},
while Sambruna et al. \cite{Sambruna} assume a specific shape of the IR
background in order to keep the TeV data compatible with the SSC model.

The corrected TeV data (Fig.~3) have a very prominent upturn above 10 TeV
which cannot be explained by SSC models.  Such an upturn may
occur naturally in some proton blazar models  \cite{Mannheim},
particularly those which predict a very flat or nearly
monoenergetic spectrum of synchrotron radiating protons
 \cite{MueckeProtheroe,Aharonian}.  However, even with these
models it appears impossible to obtain the extreme turn-up in the
TeV source spectrum inferred if the new IR data are used to
correct for propagation.

In principle the turn up could be a pile-up caused in a
pair--Compton cascade during propagation through the IR
 background \cite{ProtheroeStanev}.  However, for this to be the
 cause the intergalactic magnetic field would have to be
 extremely small otherwise the pile-up would be distributed as a
 halo \cite{CoppiAharonianVolk}, and any time-variability, such
 as the observed flaring on time scales of hours
 \cite{HEGRA501standalone} would be washed out.

\section {Discussion: What is the solution?}

Many possible solutions have been mentioned by Finkbeiner et al.
 \cite{Finkbeiner}, and here we provide a quantitative discussion
 of some of them.  The simplest solution to the problem would be
 if the recent determinations of the universal infrared
 background were an overestimate.  This is not completely ruled
 out, and they are not incompatible with star formation and
 dust-heating in the early Universe.  In a similar manner, a
 shift in the energy scale of the HEGRA data downwards by a
 factor of two would have a similar effect.  This appears rather
 extreme, and not very likely.  If both the infrared data and the
 HEGRA TeV data are confirmed, then we may be forced to consider
 more radical possibilities.

An interesting idea based on known physics was proposed by Harwit,
Protheroe and Biermann  \cite{HarwitProtheroeBiermann}.  It
considers a TeV Cherenkov event as being due to a coherent
superposition of a number of lower energy photons simultaneously
arriving at the top of the atmosphere, and masquerading as a
single TeV photon with the sum of the energies.  Such
Bose-Einstein condensates would suffer losses on a distance scale
comparable to that of the mean free path in the infrared
radiation of the individual photons of the condensate.  This idea
can be tested on the basis of extensive simulations, and by
detailed comparison of Cherenkov images of gamma ray showers from
the Crab nebula and Markarian 501.  Such analysis is underway.
\footnote{After submission of this paper a preprint by the HEGRA
Collaboration appeared (astro-ph/0006092) which renders the
possibility of a Bose-Einstein condensate explanation as being rather
unlikely.}  Even for a Bose-Einstein condensate with, on average, as
few as two arriving photons, the effect on the intergalactic
absorption of 20 TeV photons from Markarian 501 would be dramatic.  We
show in Fig.~4 the result of correcting for intergalactic absorption
for this case in an approximate way (plotted as asterisks).  In
calculating this, we assumed for simplicity that each event recorded
by the telescope consisted of a Bose-Einstein condensate of two
photons, and we then calculated the average number $N_0$ of photons of
energy $E_\gamma/2$ that would have been emitted for each pair of
photons arriving with total energy $E_\gamma$.  The energy of the
emitted bunch was then on average $N_0E_\gamma/2$, and the correction
factor used was therefore $N_0/2$.  Because of the change in
occupation number, and hence energy, of the bunch of photons during
propagation, the precise meaning of the corrected spectrum
($E_\gamma^2 F(E_\gamma)$) is ambiguous, and caution should be applied
when interpreting it.  Nevertheless, integrating the corrected
luminosity ($E_\gamma^2L(E_\gamma)$) over $\ln (E_\gamma)$ will give
the correct total emitted power.  Thus, we note that this process is a
viable mechanism for solving the problem, and has not yet been ruled
out by the observations.

Another possibility which has been suggested is violation of Lorentz
invariance \cite{ColemanGlashow,Amelino-Camelia,Kifune,Aloisio00}.
Solutions of that kind have also been invoked \cite{Sato2000} to
explain the observed high energy particles above the GZK
\cite{Gre66,Zat66} cut-off.  Amelino-Camelia et
al. \cite{Amelino-Camelia} suggested that the velocity of propagation
of energetic particles is modified {\it in vacuo} due to microscopic
quantum fluctuations occurring on scales of the order of the Planck
length.  Their dispersion relation was used by Kifune \cite{Kifune} to
show that intergalactic space will be much more transparent to TeV
photons.  We follow his treatment, in which energetic photons have
velocity $\beta_\gamma c=(1- \xi_\gamma E_\gamma/E_0)c$, and momentum
$p$ given by $p^2 c^2 =(E^2 + \xi_\gamma E_\gamma^3/E_0)$ from which
we can define an effective mass $m_\gamma$ for high energy photons by
$m_\gamma^2c^4 = -\xi_\gamma E_\gamma^3/E_0$.  Taking $\xi_\gamma=1$
and $E_0=1.2 \times 10^{19}$~GeV (Planck mass)
\cite{ColemanGlashow,Amelino-Camelia}, and using $\beta_\gamma$ and
$m_\gamma^2$ in Equation 1, we get the result shown by the thick chain
line in Figure 2.  One should be cautious, however, as Equation~1 was
derived assuming $s$ to be Lorentz invariant.  Nevertheless, our
result gives an indication of the likely effect of Lorentz Invariance
violation.  Applying our result to correct the HEGRA 1997 Markarian
501 data we obtain the spectrum shown by the diamonds in Fig.~4, and
find that indeed the absorption in intergalactic space is sufficiently
reduced to solve the problem, resulting in a spectrum that is largely
consistent with both leptonic and hadronic models.

However, it is remarkable that an acceptable source spectrum for
Markarian 501 is obtained simply by choosing the Planck mass as
the energy scale for violation of Lorentz Invariance.  Limits for
this energy scale have been obtained previously by looking for
time delays in variable sources such as Gamma Ray Bursts
\cite{Schaefer} and Markarian 421 \cite{Biller99}.  By observing
instead the energy spectrum of distant sources like Markarian 501
above 10~TeV, we may also have a sensitive tool for exploring the
energy scale of Lorentz Invariance violation.

New data on Markarian 501 and Markarian 421 relevant to the issue
discussed in this paper are soon to come as both sources are
observed to be in a high flaring state in the ongoing observation
period (May 2000).  Detecting TeV gamma-rays from a more distant
BL Lac object (e.g. $z$ =0.03--0.09) would provide for a much
needed additional constraint on the phenomenon observed at
Markarian 501.\\

\noindent {\bf \large Acknowledgements}\\
The research of RJP is funded by the Australian Research Council.
HM thanks the University of Adelaide for hospitality while this
work was carried out.


\newpage

\begin{figure}
\epsfig{file=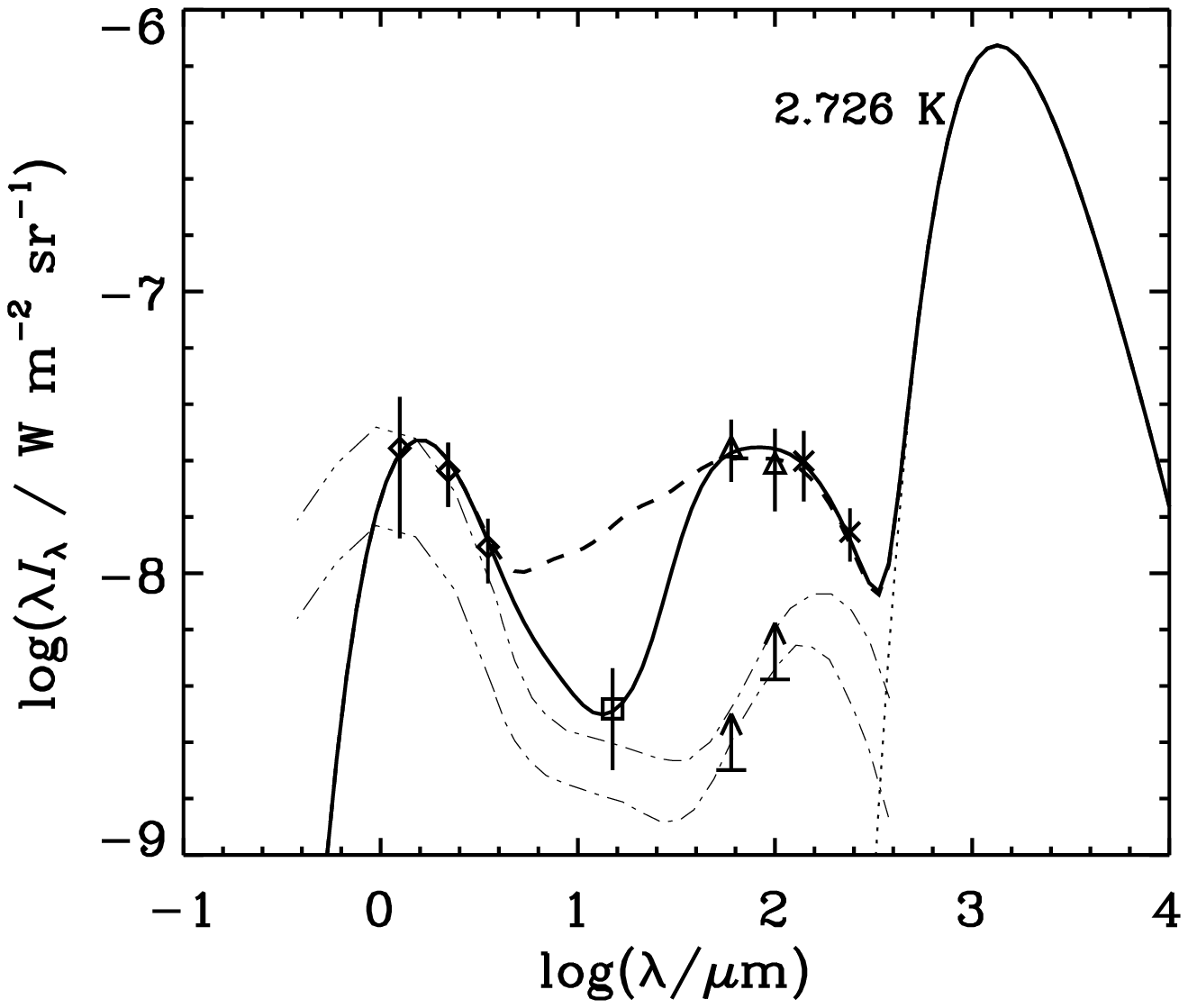, width=18cm}
\caption{Infrared background radiation field. 
 Recent determinations from DIRBE~\protect \cite{Wright}
(diamonds), ISOCAM\protect \cite{Biviano} (square),
DIRBE\protect \cite{Finkbeiner} (triangles), and
FIRAS\protect \cite{Hauser} (crosses) are compared with the models
of Malkan and Stecker\protect \cite{MalkanStecker} (dot-dash
curves; the dot-dot-dot dash curves are an extrapolation of the
Malkan and Stecker models used in
ref.~\protect \cite{BednarekProtheroe99}.  The lower limits are
from ref.~\protect \cite{WangBiermann}.  In this paper we model
the IR background by the solid curve; thick dashed curve is
allowed if the ISOCAM point is considered a lower limit.}
\end{figure}

\newpage

\begin{figure}
\epsfig{file=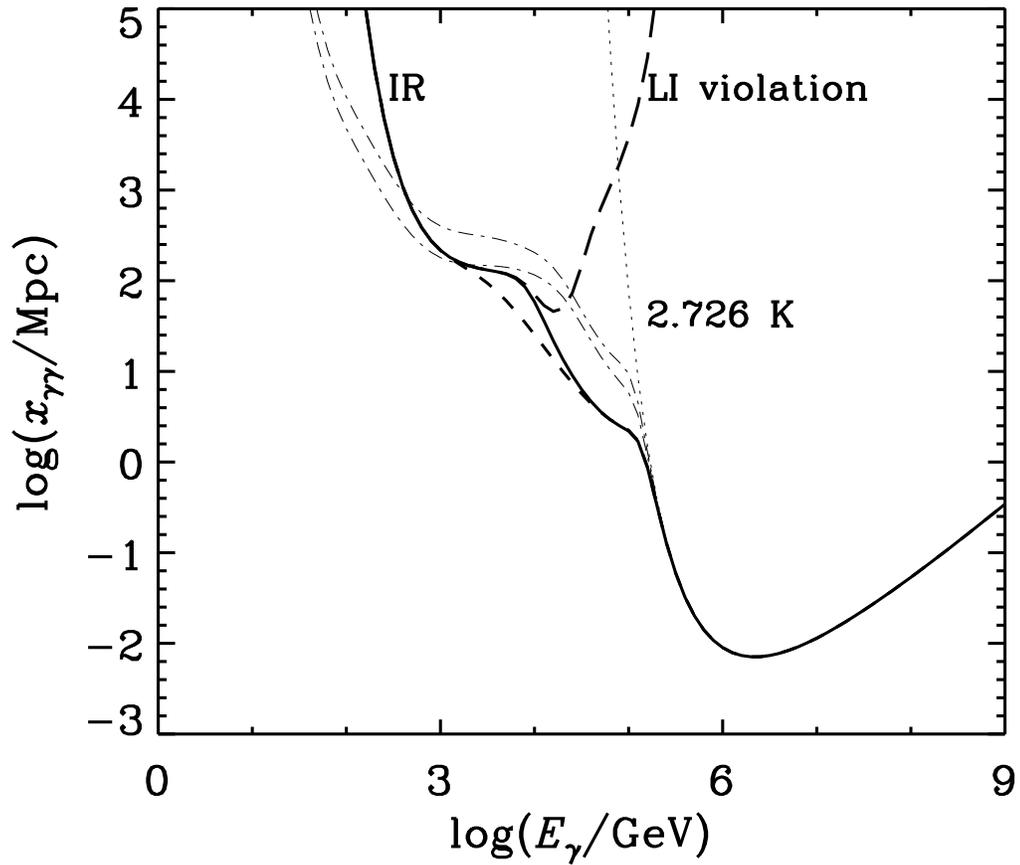, width=18cm}
\caption{Mean free path for photon-photon pair production in 
the infrared--microwave background radiation.  The curves correspond to
those in Fig.~1 except that the effect of Lorentz Invariance violation
discussed in Section 4 is shown by the long dashed curve.}
\end{figure}

\newpage

\begin{figure}
\epsfig{file=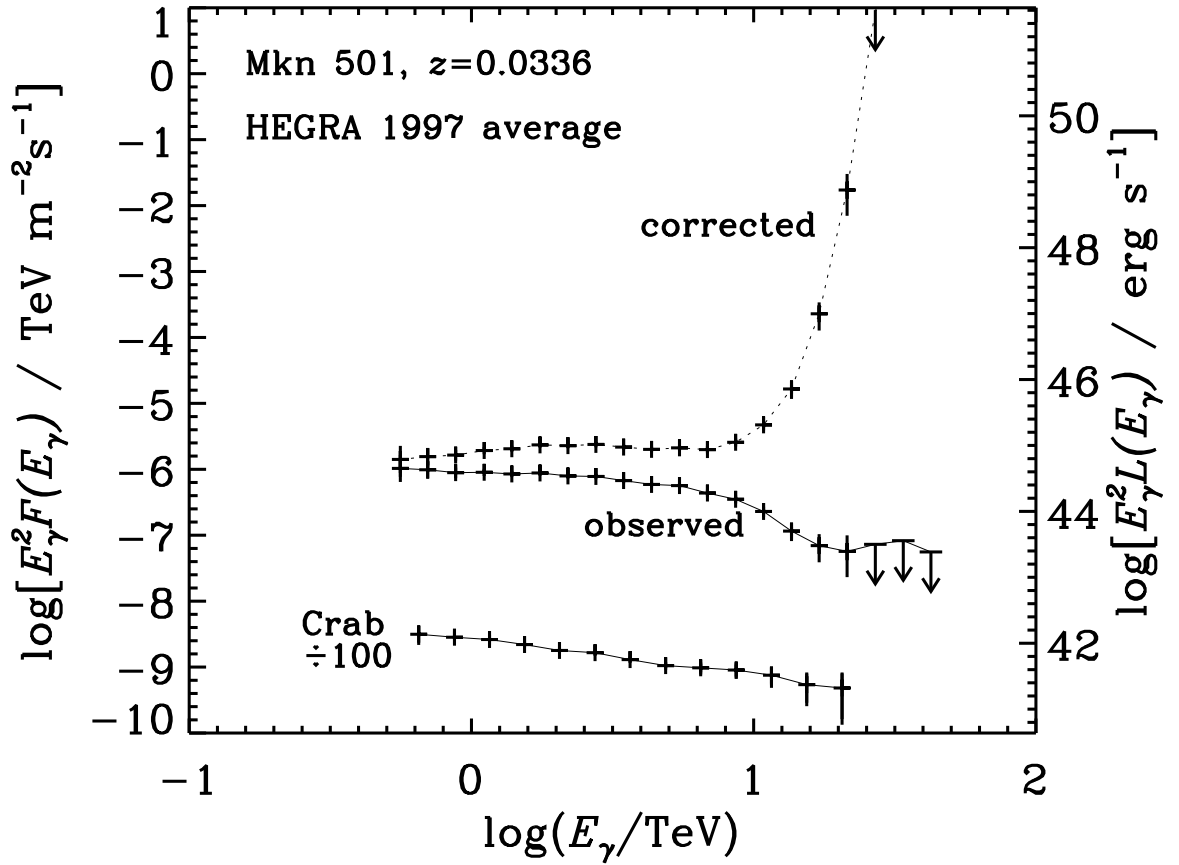, width=18cm}
\caption{The time-averaged spectrum of gamma-rays from Markarian 501 
observed in 1997\protect \cite{HEGRA97data501} is compared with
the spectrum of the Crab Nebula observed in
1997--8\protect \cite{HEGRA501standalone,HEGRAcrab}.
The spectrum of Markarian 501 after correction for absorption in
the infrared background is also shown assuming $H_0=65$
km~s$^{-1}$~Mpc$^{-1}$.  The right hand scale shows the
luminosity for Markarian 501.}
\end{figure}

\newpage

\begin{figure}
\epsfig{file=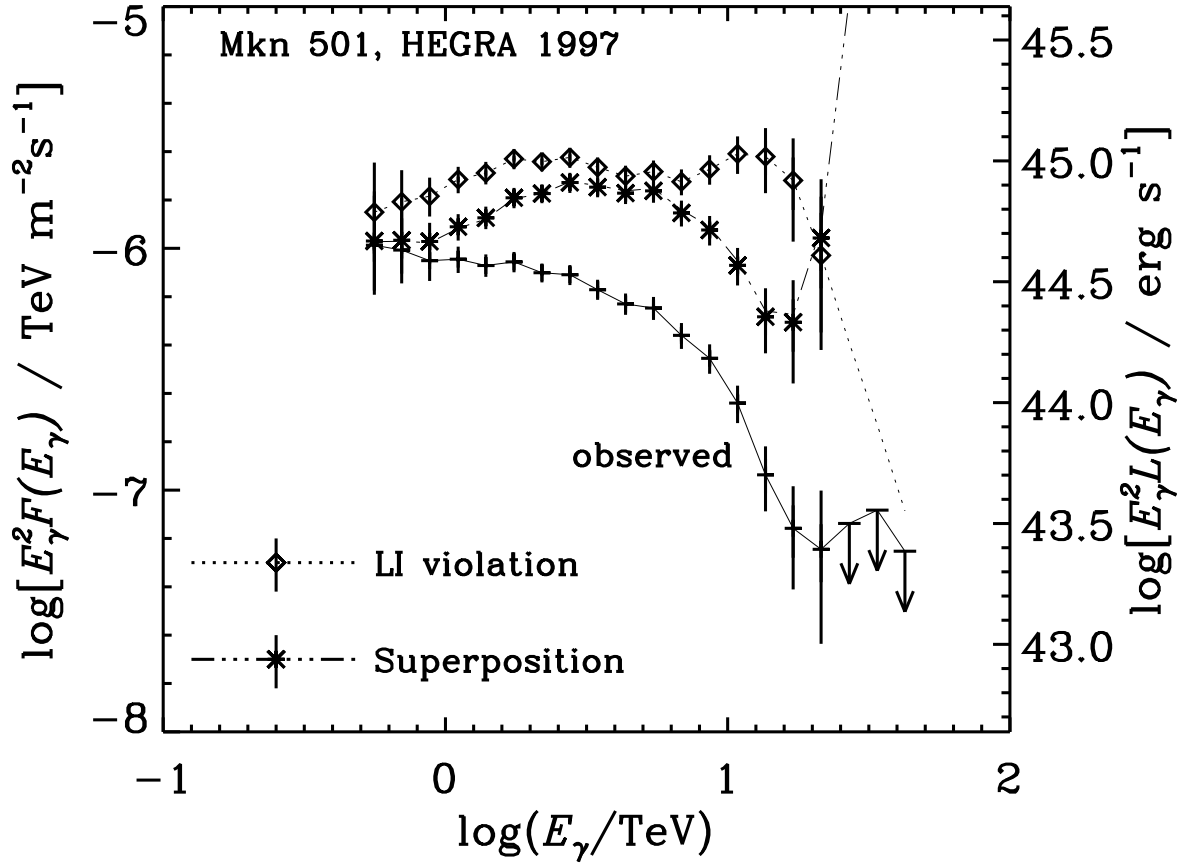, width=18cm}
\caption{The time-averaged spectrum of gamma-rays from Markarian 501 
observed by HEGRA in 1997 is plotted and compared with
the spectrum after correction assuming that each observed ``gamma
ray'' is a coherent superposition of two photons (Superposition),
or that Lorentz Invariance is violated (LI violation); upper
limits have been omitted for clarity from the ``corrected'' data. }
\end{figure}

\end{document}